\documentclass[12pt]{spieman} % 12pt font required by SPIE;
\usepackage{amsmath,amsfonts,amssymb}
\usepackage{graphicx}
\usepackage{setspace}
\usepackage{tocloft}

\title{System Design of the Newest Generation Detector Controller for ELT and new VLT Instruments}

\author[a]{Mathias Richerzhagen}
\author[a]{Matthias Seidel}
\author[a]{Leander Mehrgan}
\author[a]{Derek Ives}
\author[a]{Ralf Conzelmann}
\author[a]{Mirko Todorovic}
\author[a]{Christoph Geimer}
\affil[a]{European Southern Observatory (ESO), Karl-Schwarzschild-Str. 2, 85748 Garching bei München, Germany}

\cftpagenumbersoff{figure}
\cftpagenumbersoff{table} 
\begin{document} 
\maketitle

\begin{abstract}
A new detector controller, NGCII,  is in development for the first-generation instruments of the ELT as well as new instruments for the VLT. Building on experience with previous ESO detector controllers, a modular system based on the MicroTCA.4 industrial standard, is designed to control a variety of infrared and visible light scientific and wavefront sensor detectors. This article presents the early development stages of NGCII hardware and firmware from the decision to start an all-new design to first tests with detectors and ROICs.
\end{abstract}

% Include a list of up to six keywords after the abstract
\keywords {detector controller, NGCII, ELT, VLT, MicroTCA.4, modular system design, electronics, firmware }

\begin{spacing}{2} % use double spacing for rest of manuscript

%%%%%%%%%%%%%%%%%%%%%%%%%%%%%%%%%%%%%
%%%%%%%%%%%%%%   SOME MACROS   %%%%%%%%%%%%%%
%%%%%%%%%%%%%%%%%%%%%%%%%%%%%%%%%%%%%

\newenvironment{conditions}
  {\par\vspace{\abovedisplayskip}\noindent\begin{tabular}{ll}}
  {\end{tabular}\par\vspace{\belowdisplayskip}\vspace{\belowdisplayskip}}

%%%%%%%%%%%%%%%%%%%%%%%%%%%%%%%%%%%%%
%%%%%%%%%%%%%%   INTRO       %%%%%%%%%%%%%%%%%
%%%%%%%%%%%%%%%%%%%%%%%%%%%%%%%%%%%%%

\section{Introduction}
\label{sect:intro} % \label{} allows reference to this section
There is a long history of detector controller development at ESO. Starting with IRACE\cite{Meyer1998} and FIERA\cite{Beletic1998} and continuing with NGC\cite{Mehrgan2013} there has been the capability to control visible and infrared scientific detectors using in-house developed controller hardware and software adapted to ESO telescope infrastructure. For ELT instruments, development of new VLT instruments and the upgrade of older VLT instruments it was decided to develop a new detector controller NGCII in 2021. This paper reports on fundamental design decisions made to arrive at the system design of the NGCII detector controller as well as initial performance figures from the implemented design. 

While commercial readout solutions do exist for all of the detector types in use at ESO, no one controller could control them all or meet the other requirements for compatibility with ESO infrastructure.

\begin{figure}[ht]
\centering
\includegraphics[width=\textwidth]{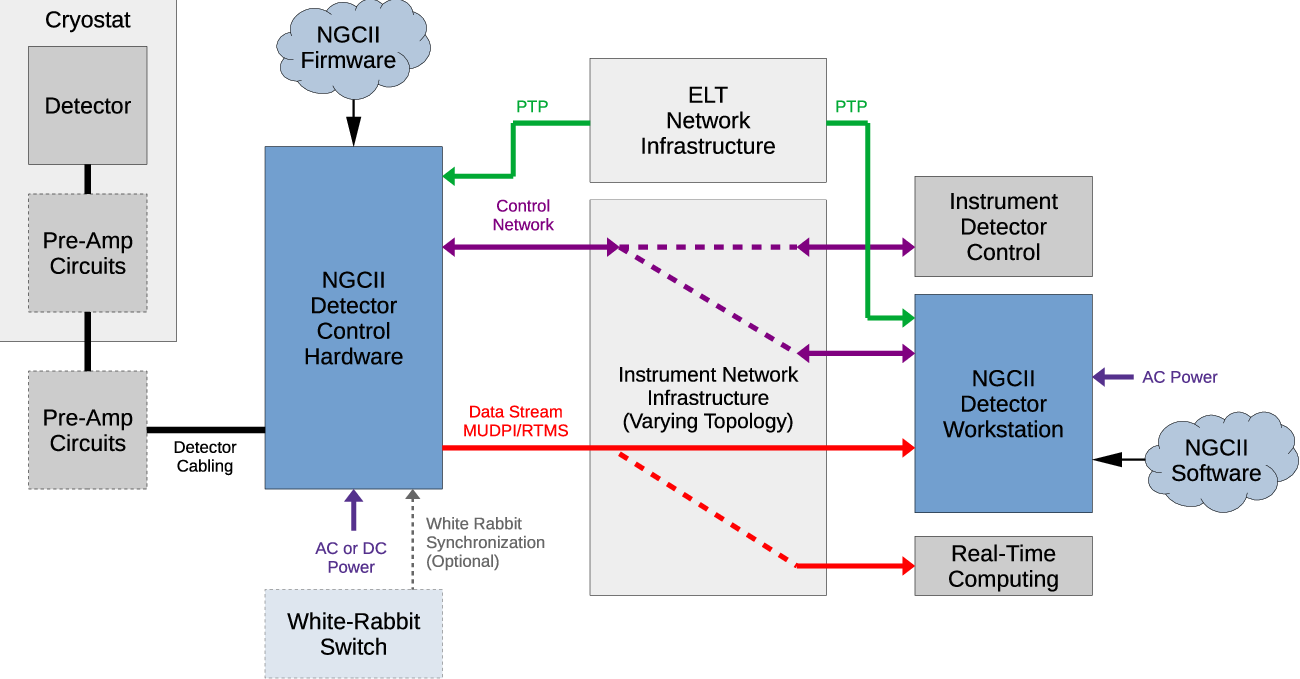}
\caption{System Block Diagram}
\label{pic:sys_bd}
\end{figure}

Figure \ref{pic:sys_bd} shows the top level system block diagram of the detector controller as covered by this paper. The detector and its cryogenic and/or warm preamp circuits are interfaced by the detector control hardware running NGCII firmware which is the core component of the detector controller. Through three separate networks the detector control hardware interfaces to the telescope infrastructure, instrument detector control systems and the detector workstation computer running the detector control software package.

With the detector control software package continuously developed from NGC and the detector workstation being a standard commercial-of-the-shelf server computer, the focus lies on the detector control hardware and associated firmware.

\section{Method and Requirements}

Requirements for a new detector controller were collected, considering the detectors required by the ELT first generation instruments HARMONI\cite{Thatte2020}, METIS\cite{Brandl2014}, MICADO\cite{Davies2016} as well as foreseeable new instruments for the VLT like MAVIS\cite{McDermid2020}, CUBES\cite{Zanutta2022}, FORS-Upgrade\cite{Cirami2020}. 

As shown in chapter \ref{sec:detectors} additional detector types are considered in the road map and the system architecture is kept as generic as possible to support as many future detector systems as possible.

\subsection{Development Goal}
\label{sec:goal}
It is not expected to achieve major direct performance gains over NGC regarding detector readout since detector system performance in current systems is already limited by the detector itself or cryogenic electronics. The goal therefore is to at least meet the detector readout performance of NGC for common detectors while making improvements for:
\begin{itemize}
\item{Operational Reliability}
\item{Stability over Change of Environmental Conditions}
\item{Features for Future Read-Out Modes}
\item{Read-out Speed}
\end{itemize}

The importance of the requirement to be robust against the change of environmental conditions was highlighted by NASA\cite{Rauscher2013} who performed principle component analysis (PCA) measurements on a HAWAII-2RG infrared detector finding that minor variations on some detector bias voltages have an impact on detector performance. Some PCA measurements were repeated on the NGC detector controller and the system was found to be affected as well\cite{George2018,George2018_2}. When deriving circuit requirements from overall operational requirements, care is taken to specify easily verifiable requirements based on principle components. For example, for a voltage source, there are separate requirements for static error, drift over ambient temperature and random noise.

\subsection{Rationale for Development of a New Controller vs. Continued Support of NGC}
The decision to develop a new detector controller versus continued support of NGC was made after evaluating the effort required to bring NGC to a state where it would be suitable for operation with ELT infrastructure. Several technical issues were identified that would have required major changes to NGC. 

\subsubsection{Communication Interfaces}
NGC uses a custom communication protocol over fibers for image transfer from the detector control hardware to the NGC equivalent of the detector workstation, as well as command and control of the detector control hardware. This is incompatible with ELT operational requirements flowed down to the common requirements for ELT instruments, and thus NGCII, that mandate the use of standard Ethernet protocols. It was considered to upgrade NGC to use standard network protocols but found that an upgrade to the necessary high-bandwidth network standard for image data transfer was not feasible.

The required image data interface bandwidth is driven by the GEOSNAP detector with its 2048x2048 pixel array, 14bit ADC and nominal 85fps frame rate, yielding a raw data rate of 5Gbps. Higher frame rates may be possible pending tests. This is closely followed by the HAWAII-2RG detector with a 2048x2048 pixel array, up to 33fps frame rate and an 18bit ADC where it is desired to read two detectors from the same controller yielding a raw data rate of 5Gbps as well. As a baseline, a 10GBASE-LR interface is selected with possible upgrade paths to 25GBASE-LR or 100GBASE-LR4 already considered early in the design, to be implemented once the need arises and the required infrastructure in the telescope and instruments becomes available.

Additionally, the ELT infrastructure relies on the precision time protocol (PTP) for synchronization of telescope subsystem, an interface not available on NGC.

\subsubsection{Internal Backplane Bandwidth}
With increasing data rate requirements of detectors, an upgrade of the controller-internal backplane was required to aggregate data from multiple modules for transmission over the communication interfaces. NGC uses a daisy chain approach for modules, which is not feasible to further upgrade for more bandwidth with reasonable effort.

\subsubsection{Component Obsolescence}
With the design of NGC being 10 years old in 2021 the frequency of component obsolescence was increasing. A re-design of each NGC module would have been required to replace obsolete components and ensure maintainability for another 20 years.

\subsection{Form Factor}
We decided to base NGCII on a 19” rack mountable enclosure, the same as NGC. Due to technological progress in the last decade, our analysis also indicated that it was possible to integrate the previously external power supply unit into the detector controller itself, further decreasing the space required for the detector control electronics. During conceptual design, an analysis showed that it would not be possible to achieve a higher packing density than NGC for every possible detector configuration, but that a realistic design goal would be to aim for a smaller form factor for most configurations while allowing for some more space for others. The design goal set out in the beginning of the project was to drive a simple detector like the HAWAII-2RG or CCD231-84 from a single height unit (HU) on average while requiring 2HU for more complex detectors.

Anticipating the higher communication bandwidth requirements, more intensive local processing and general higher circuit complexity to meet functional requirements, it was found that it would not be feasible to attempt a significantly lower power consumption than NGC even considering availability of higher efficiency power supplies. 

\subsection{Supported Detectors}
\label{sec:detectors}
NGCII is intended be a modular detector controller designed to control any, infrared or visible light, scientific or wavefront sensor detector used in ground-based astronomy. Detector support is rolled out in multiple stages with detectors in Table \ref{tab:detectors1} included in a first development milestone and detectors in Table \ref{tab:detectors2} considered during conceptual design with no hardware development ongoing. Any other detectors may be added into the scope of NGCII at a later date as a follow-on project.

\begin{table}[ht]
\caption{List of Detectors, Milestone 1} 
\label{tab:detectors1}
\small
\begin{center}
\begin{tabular}{|l|l|l|l|} %% this creates two columns
\hline
\rule[-1ex]{0pt}{3.5ex} Manufacturer &Detector & Ref. & Type  \\ \hline \hline
\rule[-1ex]{0pt}{3.5ex} Teledyne & HAWAII 2RG & Ref.~[\citenum{Loose2003}] & CMOS \\ \hline
\rule[-1ex]{0pt}{3.5ex} Teledyne & HAWAII 4RG & Ref.~[\citenum{Hodapp2022}] & CMOS \\ \hline
\rule[-1ex]{0pt}{3.5ex} Leonardo & SAPHIRA & Ref.~[\citenum{Finger2014}]  & eAPD CMOS \\ \hline 
\rule[-1ex]{0pt}{3.5ex} Teledyne E2V & CCD231-84 & Ref.~[\citenum{CCD231}] & CCD \\ \hline
\rule[-1ex]{0pt}{3.5ex} Teledyne E2V & CCD290-99 & Ref.~[\citenum{CCD290}] & CCD \\ \hline
\rule[-1ex]{0pt}{3.5ex} Teledyne Imaging Sensors & GEOSNAP & Ref.~[\citenum{Leisenring2023}] & Fully Digital ROIC \\ \hline
\end{tabular}
\end{center}
\end{table}

\begin{table}[ht]
\caption{List of Detectors, Roadmap} 
\label{tab:detectors2}
\small
\begin{center}
\begin{tabular}{|l|l|l|l|} %% this creates two columns
\hline
\rule[-1ex]{0pt}{3.5ex} Manufacturer & Detector & Ref. & Type  \\ \hline \hline
\rule[-1ex]{0pt}{3.5ex} Leonardo & Large SAPHIRA & Ref.~[\citenum{Hall2016}]   &  eAPD CMOS \\ \hline 
\rule[-1ex]{0pt}{3.5ex} Lawrence Berkeley National Laboratory & LBNL CCD & Ref.~[\citenum{Holland2002}] & Fully Depleted CCD \\ \hline
\rule[-1ex]{0pt}{3.5ex} SOFRADIR & ALFA & Ref.~[\citenum{Fieque2018}] & CMOS \\ \hline
\rule[-1ex]{0pt}{3.5ex} Raytheon & AQUARIUS & -  & - \\ \hline
\rule[-1ex]{0pt}{3.5ex} Raytheon & VIRGO & - &  -\\ \hline
\rule[-1ex]{0pt}{3.5ex} Semiconductor Technology Associates & STA CCD & -  & - \\ \hline
\rule[-1ex]{0pt}{3.5ex} Gpixel & GSENSE & - & - \\ \hline
\end{tabular}
\end{center}
\end{table}

%-----------------------------------
%    MTCA 
%-----------------------------------
\section{Platform Selection}
Previous detector control systems at ESO were developed as modular systems, which, specifically with NGC, has been a successful approach. Therefore, an early decision in development was to develop a modular system again. The segment of small form-factor wavefront sensing cameras is already covered by the ALICE and LISA cameras\cite{Marchetti2022} so the unavoidable overhead of a modular system regarding space, power consumption and complexity is accepted.

Whilst not implementing a standard-compliant VMEbus, NGC was loosely based on the VME mechanical form factor. Since VME is largely considered obsolete at the time of NGCII design, a different, newer, pre-existing modular rack hardware standard was searched and found. A complete from-scratch development was considered as well but not pursued any further due to the amount of development resources required.

Very early during the conceptual design phase it was chosen to base NGCII on the MicroTCA.4\cite{mtca4} modular standard. When surveying modular rack equipment standards under active support this was chosen over the competing standard PXI\cite{pxi2020} for its wide use in the particle and high energy physics community, for example at the European XFEL\cite{Branlard2013}, DESY\cite{Walter2019}, CERN\cite{Bobillier2016}, and the European Spallation Source ESS\cite{Chicken2023}. 

\subsection{MicroTCA.4}
The MicroTCA family of standards is derived from the telecoms industry standard ATCA. MicroTCA.4 extends the base standard by rear-transition modules as well as precision timing options for use in physics applications. MicroTCA.4 is hosted by the PCI Industrial Computer Manufacturers Group (PICMG). This section gives a short introduction into MicroTCA.4 in the context of NGCII.

\subsubsection{Advanced Mezzanine Cards}
\begin{figure}[ht]
\centering
\includegraphics[width=0.7\textwidth]{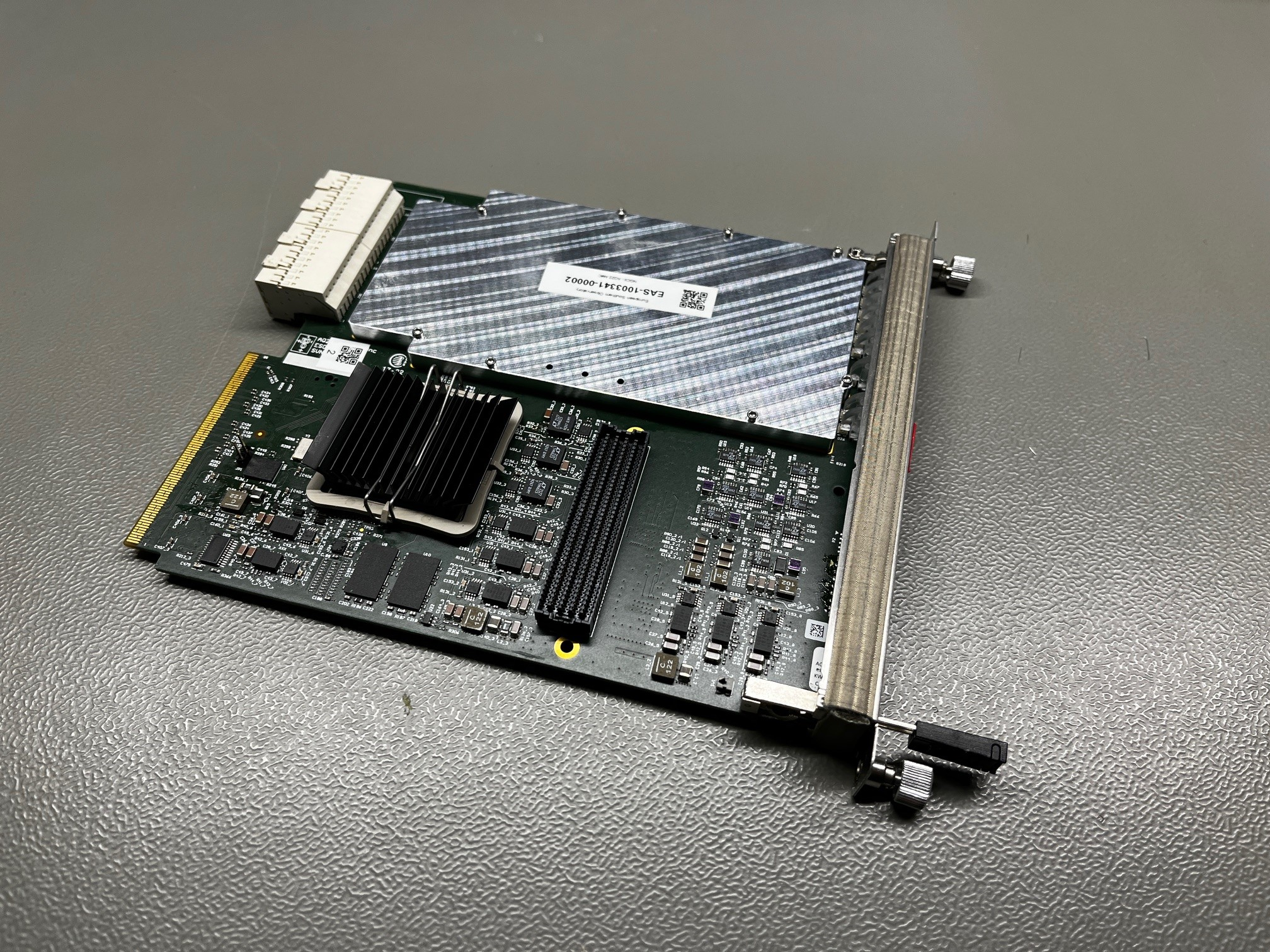}
\caption{Picture of AQ22 AMC Module}
\label{pic:aq22}
\end{figure}

The advanced mezzanine card (AMC) is the core module type of a MicroTCA system. It interfaces the backplane through the zone 1 connector which provides power, synchronization signals and high-bandwidth, packet-based communication. The AMC module typically contains a CPU, FPGA or system on chip (SoC) as well as user-specific circuits. A picture of an AMC module is shown in Figure \ref{pic:aq22}. 

\subsubsection{Rear Transition Modules}
\label{sec:rtm}
\begin{figure}[ht]
\centering
\includegraphics[width=0.9\textwidth]{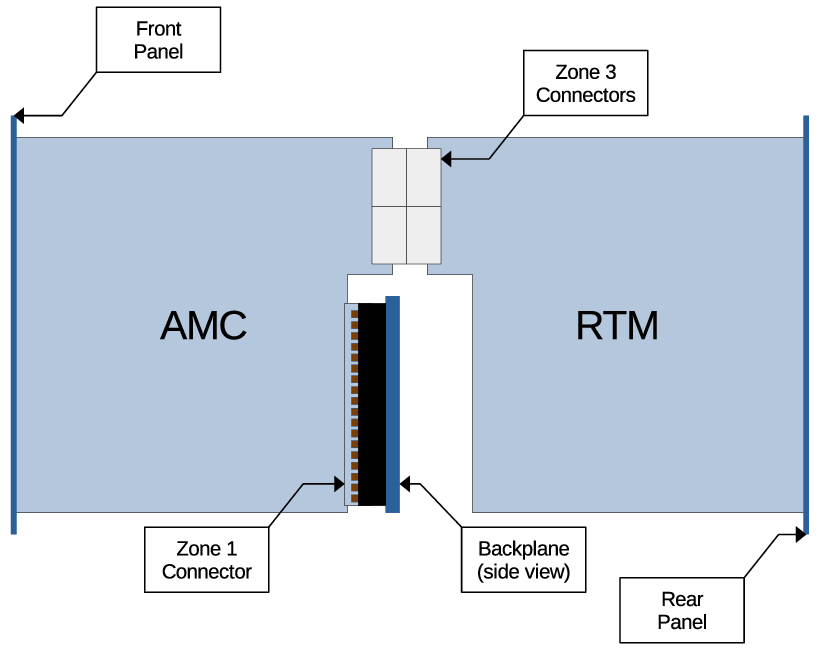}
\caption{AMC/RTM Connection}
\label{pic:rtm_amc}
\end{figure}

The rear transition module (RTM) is an extension of the AMC towards the rear of the backplane. It has no direct connection to the backplane but only connects to the AMC module through the zone 3 connector. Figure \ref{pic:rtm_amc} shows the connectivity between AMC, RTM and backplane.
The connectors and signals of the zone 3 connector are not defined by the standard, but left open for the user to define. In typical system configurations, ZD-type metric backplane connectors are used\cite{zone3_digital,zone3_analog} carrying a combination of:
\begin{itemize}
\item{High speed current mode logic (CML) transceivers signals}
\item{Low-voltage differential signaling (LVDS) fixed function I/O signals}
\item{LVDS user I/O signals}
\item{Fully differential analog signals}
\end{itemize}
For NGCII it was decided, as much as practically possible, to keep detector specific circuits on the RTM.

\subsubsection{Shelves}
In MicroTCA terminology a shelf is typically a 19'' sub rack carrying multiple AMC and RTM cards as well as the power supply unit (PSU), MicroTCA controller hub (MCH), a backplane, and cooling unit (CU). As a default configuration the MicroTCA.4 standard\cite{mtca4} assumes a 12-slot configuration with redundant MCH and power supply but smaller shelves with four to seven slots and no redundancy are also commonly available \cite{nativeR2,vt814}.

\subsubsection{Backplane Topology}
The MicroTCA.4 standard\cite{mtca4} defines recommendations for backplane connectivity but leaves implementation to the equipment manufacturers.
\begin{figure}[ht]
\centering
\includegraphics[width=0.9\textwidth]{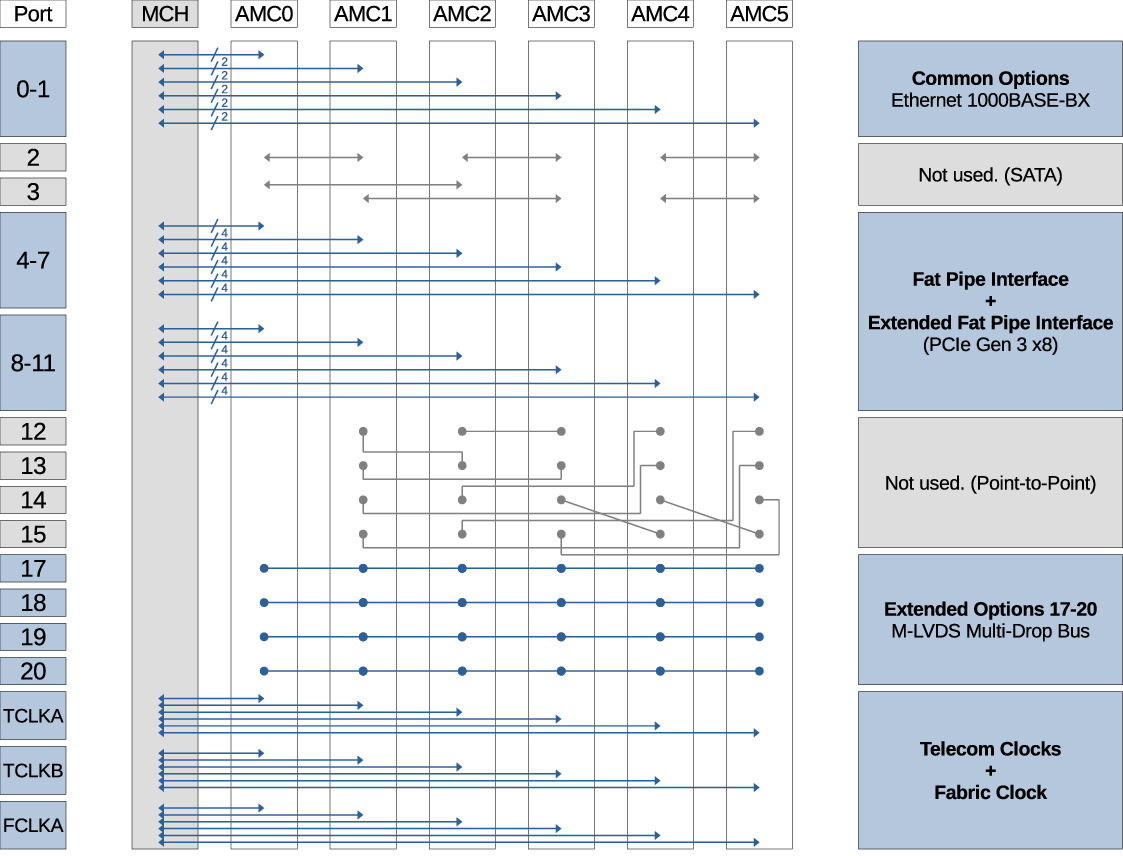}
\caption{Simplified Backplane Topology}
\label{pic:backplane}
\end{figure}
A full description of the MicroTCA backplane topology with all options would go beyond the scope of this work so Figure \ref{pic:backplane} shows a simplified diagram of a MicroTCA.4 backplane for a 6-slot shelf\cite{nativeR2} with only those options shown that are commonly available in commercial MicroTCA shelves. Connectivity that is used in other MicroTCA installations but not in NGCII is grayed out. In the context of MicroTCA, a port refers to a pair (nominally RX and TX) of differential signals available on the AMC connector.

\subsubsection{Power Supply Unit (PSU)}
MicroTCA uses a $3.3V$ management power supply and $12V$ payload power supply per slot, generated by the power supply unit. Power use is monitored and managed individually per slot.  RTMs are powered through their associated AMCs. There are several options available for system power input. Most widely available is $230V_{AC}$,$50Hz$ or $110V_{AC}$,$60Hz$ but operation from a $+24V_{DC}$ or $-48V_{DC}$ bus is also possible by selecting the appropriate power supply unit. In the context of ESO detector controllers, an integrated power supply is a departure from the external 19'' power supply used on NGC.

\subsubsection{MicroTCA Controller Hub (MCH)}
The MicroTCA controller hub (MCH) is the central communication and management hub in a MicroTCA system. It provides switched Gigabit Ethernet to all slots as a basic communication layer and the intelligent platform management interface (IPMI) as a management layer. Clock signals are distributed through the TCLK and FCLK clock fabrics. Clocks can be generated on the MCH and distributed to AMCs or sourced from one AMC and distributed to others. Finally the MCH acts as a fat-pipe switch for the high bandwidth packet based communication interface with the standard suggesting using PCI Express, Serial Rapid I/O (SRIO)\footnote{SRIO is a relic from the ATCA legacy of MicroTCA. There are very few products available that support SRIO at the time of writing.} or 10GBit Ethernet. In case of NGCII, PCIe Gen3 fat-pipe switches are selected.

\subsubsection{MicroTCA Management}
MicroTCA implements a management plane\cite{mtca4} for module identification, electronic keying, thermal management and sensor readout via an intelligent platform management interface (IPMI) bus. If implemented on every module, as mandated by the standard, it provides full system diagnostics at a low level and which is independent of both NGCII firmware and software.

%-----------------------------------
%      General Architecture 
%-----------------------------------
\section{NGCII Architecture}
The design of NGCII yields a mixture of commercial components and in-house developed modules making up the detector control hardware. During conceptual design, the required functionality was distributed over multiple AMC and RTM type modules. For each module a make-or-buy decision was made to evaluate if a module is best bought off-the-shelf or developed in house.

\subsection{Commercial Modules}
As far as possible, commercial-off-the-shelf modules were selected for NGCII components to reduce development effort with the limited resources available. Due to the widespread use of MicroTCA.4 in the particle physics community, there is, at present, an ecosystem of suppliers that design and manufacture MicroTCA compliant hardware.

\subsubsection{MicroTCA.4 Shelf with Power Supply and MCH}
\label{sec:shelf}
For the shelf, a market study was performed optimizing the density of controlled detectors per rack height-unit. As shown in section \ref{sec:sysconfig} a typical detector requires three to five modules to drive it with the option to share an acquisition module between two detectors in some cases. The result of the study was that a 2HU shelf with five or six AMC and RTM slots yields an optimum of one detector per 2HU or, in some high-density configurations, one detector per 1HU.

\begin{figure}[ht]
\centering
\includegraphics[width=0.7\textwidth]{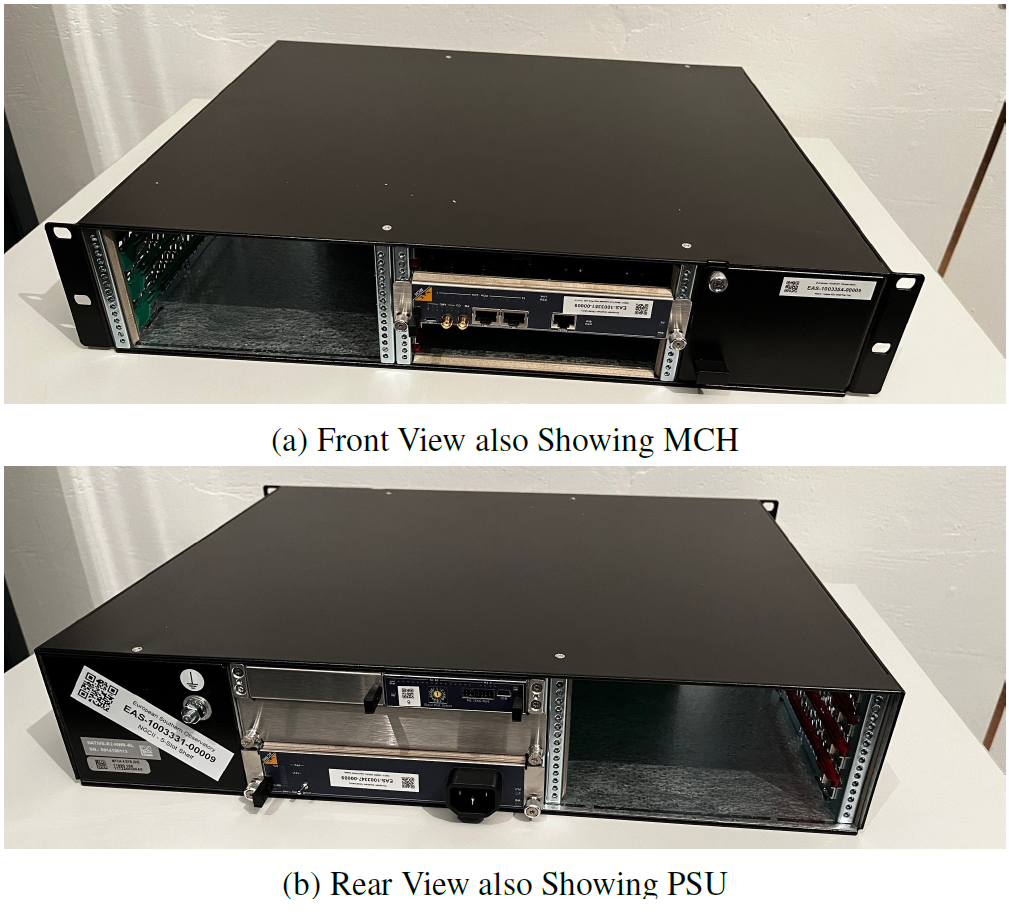}
\caption{Pictures of 5-Slot Shelf}
\label{pic:shelves}
\end{figure}

A call for tender was won by the N.A.T. Native R2 \cite{nativeR2} 5-slot, 2HU shelf shown in Figure \ref{pic:shelves}. It includes an MCH and a PSU from the same manufacturer. This shelf is used as the baseline for most configurations while another competition for a 6-slot, 2HU shelf is ongoing to cover the remaining system configurations that cannot be implemented using the 5-slot shelf.

\subsubsection{Host Controller SoC Module}
\label{sec:host-controller}
A host controller AMC module is required for central control of the detector control hardware, data aggregation and main host for the NGCII firmware. On the ALICE and LISA wavefront sensor cameras\cite{Marchetti2022} this role is fulfilled by a Xilinx/AMD Zynq Ultrascale+ SoC with ARM CPU cores and a programmable logic FPGA fabric. A SoC from the same family is selected for NGCII. In the NGCII context the module acts as root complex on the shelf internal PCIe bus over which it controls all other AMC modules and aggregates data. The Ethernet interfaces for communication are also located on this module.

A call for tender was won by the IOxOS IFC-1414. The module contains the required SoC, four  small form-factor pluggable (SFP+) transceiver slots for communication interfaces and a digital LVDS/CML zone 3 interface to host a detector control RTM.

\subsubsection{Auxiliary Controller Module}
\label{sec:aux-controller}
The host controller described in section \ref{sec:host-controller} is used for most computing tasks in the system and hosts one RTM with a digital zone 3 profile, for example a clock and/or bias module. As shown in section \ref{sec:sysconfig} several system configurations require more than one RTM to drive a detector, for example, in case of CCDs where clock and bias modules are separate RTMs or SAPHIRA where and additional APD Bias RTM is required. For those configurations the need for a lightweight FPGA module with a Xilinx/AMD Spartan 7 or Artix 7 FPGA and digital zone 3 profile was identified.

For this module, a call for tender was won by the AIES MFMC. The module contain a Xilinx/AMD Artix 7 FPGA, the required LVDS/CML zone 3 profile and two FPGA mezzanine card (FMC) slots that can be used for future expansions.

\subsection{ESO Specific Modules}
Modules that were not commercially available at the time of conceptual design are designed in-house at ESO. While the particle physics community has commercialized many modules required to drive particle physics experiments like timing cards, safety interlock systems or RF metrology cards\cite{Fenner2023} no such hardware exists for astronomical telescopes. Therefore all detector-control specific modules are developed in-house at ESO, building on the well-proven base of NGC.

\subsubsection{General Acquisition AMC (AQ22 AMC)}
\label{sec:aq22}
For digitization of analog video signal an AMC module containing ADCs is required. At the time of design no commercial off-the-shelf module meeting ESO requirements was available so an in-house development was started.

The requirements for the ADC have been identified as minimum 16-bit resolution with maximum $1 LSB_{16}$ non-linearity. Furthermore for digitization of the detector video signals, an ADC that allows determining the sample time by an external trigger signal is required, which excludes many ADC topologies. Finally the LTC2387-18\cite{LTC2387} ADC was selected which features 18-bit resolution at $15 Msps$ with less than $1 LSB_{18}$ typical non-linearity error. The sample-and-hold circuit of the ADC can be controlled by an external digital signal, allowing hardware synchronization of sample timing.

For determining the optimum number of video channels per module, an analysis was performed taking into account the available board space and number of video channels required by detectors. With HAWAII-2RG and HAWAII-4RG detectors having 33/66 video outputs\footnote{Including reference channels and window output channels.} respectively, it was decided to build the acquisition modules on multiples of 11 channels which leads to 22 video channels per module. In section \ref{sec:sysconfig} the system configurations are shown.

\subsubsection{Detector Control RTMs}
Besides the AQ22 acquisition AMC, all other detector specific circuits are implemented as rear-transition-modules (RTMs). For the first milestone covering detectors listed in Table  \ref{tab:detectors1} the following seven RTMs are developed.
\begin{table}[ht]
\caption{List of RTMs} 
\label{tab:rtms}
\small
\begin{center}
\begin{tabular}{|l|l|l|} %% this creates two columns
\hline
\rule[-1ex]{0pt}{3.5ex} Name & Zone 3 & Features \\ \hline \hline
\rule[-1ex]{0pt}{3.5ex} CMOS C20B20 RTM & Digital LVDS & 20x CMOS Bias, 20x CMOS Clock \\ \hline 
\rule[-1ex]{0pt}{3.5ex} CMOS AQ22 RTM & Analog & 22x CMOS Video Channel, 1x Preamp Supply \\ \hline 
\rule[-1ex]{0pt}{3.5ex} APD Bias RTM & Digital LVDS & 2x APD Bias Channel \\ \hline 
\rule[-1ex]{0pt}{3.5ex} CCD B24 RTM & Digital LVDS & 24x CCD Bias Channel \\ \hline 
\rule[-1ex]{0pt}{3.5ex} CCD C24 RTM & Digital LVDS & 24x CCD Clock Channel \\ \hline 
\rule[-1ex]{0pt}{3.5ex} CCD AQ8 RTM & Analog & 8x CCD Video Channel, 1x Preamp Supply \\ \hline 
\rule[-1ex]{0pt}{3.5ex} GEOSNAP RTM & Digital LVDS/CML & GEOSNAP Specific Circuits \\ \hline 
\end{tabular}
\end{center}
\end{table} 

\begin{figure}[ht]
\centering
\includegraphics[width=0.9\textwidth]{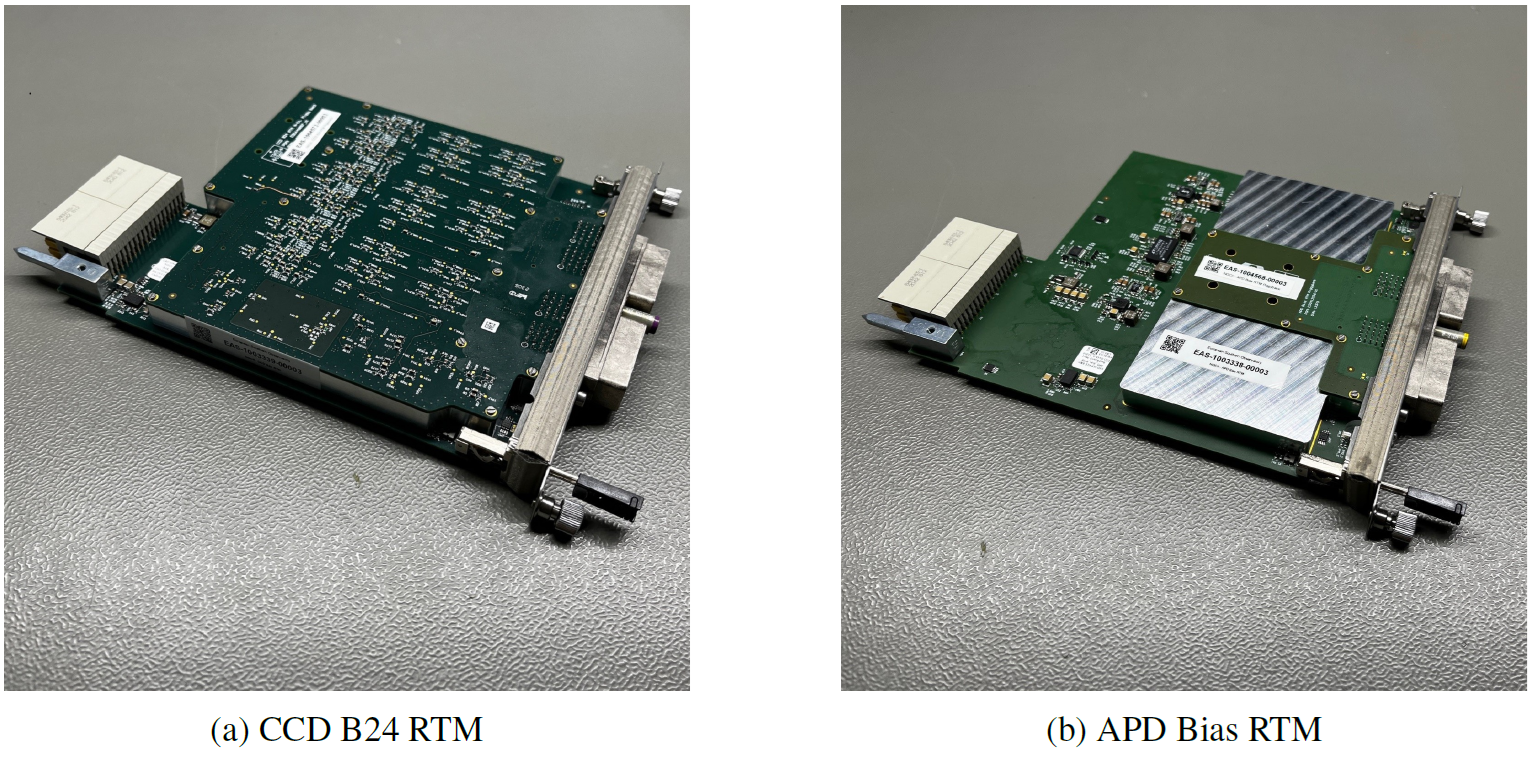}
\caption{Examples for ESO-Developed Modules}
\label{pic:modules}
\end{figure}

The two video RTMs with an analog zone 3 profile interface the ESO-designed AQ22 AMC described in section \ref{sec:aq22}, modules with digital zone 3 profile interface the commercial FPGA modules described in sections \ref{sec:host-controller} and \ref{sec:aux-controller}. Figure \ref{pic:modules} shows pictures of two examples for ESO-developed RTM modules.

\subsection{System Configurations}
\label{sec:sysconfig}
Examples of system configurations are shown in this section. The list of configurations is not complete but shows selected examples that highlight the design of NGCII. All configurations are based on the MicroTCA.4 shelf, PSU and MCH described in section \ref{sec:shelf}.

\subsection{CMOS Detector Configurations}
Configurations are available for the HAWAII-2RG, HAWAII-4RG and SAPHIRA detectors. The most basic configuration is a single detector per shelf as shown in Table \ref{tab:single_cmos}. The APD Bias RTM is only used for the SAPHIRA detector and is not required for HAWAII-2RG systems. There are 11 unused video channels in this configuration which is a trade-off taken for being able to implement higher density configurations.
\begin{table}[ht]
\caption{Single CMOS Detector} 
\label{tab:single_cmos}
\small
\begin{center}
\begin{tabular}{|l|l|l|l|l|} %% this creates two columns
\hline
\rule[-1ex]{0pt}{3.5ex} Slot & AMC Module & Zone 3 & RTM Module & Comment \\ \hline \hline
\rule[-1ex]{0pt}{3.5ex} 1 & IOxOS IFC1414 & Digital LVDS & CMOS C20B20 RTM & \\ \hline 
\rule[-1ex]{0pt}{3.5ex} 2 & AQ22 AMC & Analog & CMOS AQ22 RTM & \\ \hline 
\rule[-1ex]{0pt}{3.5ex} 3 & AQ22 AMC & Analog & CMOS AQ22 RTM &\\ \hline 
\rule[-1ex]{0pt}{3.5ex} 4 & AIES MFMC & Digital LVDS & APD Bias RTM & Only for SAPHIRA\\ \hline 
\end{tabular}
\end{center}
\end{table} 

To achieve a higher density of detectors controlled, one detector per 1HU instead of one detector per 2HU, it is possible to operate two CMOS detectors from one shelf. Since one video module is shared between detectors A and B it is recommended to only do this if both detectors are read synchronously, for example in a detector mosaic. For the SAPHIRA configuration a 6-slot shelf is required which will be available in the future but for HAWAII-2RG detectors it is possible to implement the dual detector in a standard 5-slot shelf. The configuration is shown in Table \ref{tab:dual_cmos}.
\begin{table}[ht]
\caption{Dual CMOS Detector} 
\label{tab:dual_cmos}
\small
\begin{center}
\begin{tabular}{|l|l|l|l|l|l|} %% this creates two columns
\hline
\rule[-1ex]{0pt}{3.5ex} Slot & AMC Module & Zone 3 & RTM Module & Det. & Comment \\ \hline \hline
\rule[-1ex]{0pt}{3.5ex} 1 & IOxOS IFC1414 & Digital LVDS & CMOS C20B20 RTM & A & \\ \hline 
\rule[-1ex]{0pt}{3.5ex} 2 & AQ22 AMC & Analog & CMOS AQ22 RTM & A &\\ \hline 
\rule[-1ex]{0pt}{3.5ex} 3 & AQ22 AMC & Analog & CMOS AQ22 RTM & A/B &\\ \hline 
\rule[-1ex]{0pt}{3.5ex} 4 & AQ22 AMC & Analog & CMOS AQ22 RTM & B &\\ \hline 
\rule[-1ex]{0pt}{3.5ex} 5 & AIES MFMC & Digital LVDS & CMOS C20B20 RTM & B & \\ \hline 
\rule[-1ex]{0pt}{3.5ex} 6 & AIES MFMC & Digital LVDS & APD Bias RTM & A/B & Only for SAPHIRA\\ \hline 
\end{tabular}
\end{center}
\end{table} 

For the HAWAII-4RG and Large SAPHIRA it is required to have 64 to 66 video channels available. The configuration is shown in Table \ref{tab:large_cmos}.
\begin{table}[ht]
\caption{Large CMOS Detector} 
\label{tab:large_cmos}
\small
\begin{center}
\begin{tabular}{|l|l|l|l|l|} %% this creates two columns
\hline
\rule[-1ex]{0pt}{3.5ex} Slot & AMC Module & Zone 3 & RTM Module & Comment \\ \hline \hline
\rule[-1ex]{0pt}{3.5ex} 1 & IOxOS IFC1414 & Digital LVDS & CMOS C20B20 RTM &  \\ \hline 
\rule[-1ex]{0pt}{3.5ex} 2 & AQ22 AMC & Analog & CMOS AQ22 RTM & \\ \hline 
\rule[-1ex]{0pt}{3.5ex} 3 & AQ22 AMC & Analog & CMOS AQ22 RTM & \\ \hline 
\rule[-1ex]{0pt}{3.5ex} 4 & AQ22 AMC & Analog & CMOS AQ22 RTM & \\ \hline 
\rule[-1ex]{0pt}{3.5ex} 5 & AIES MFMC & Digital LVDS & APD Bias RTM &  Only for SAPHIRA\\ \hline 
\end{tabular}
\end{center}
\end{table} 

\subsection{CCD Detector Configurations}
CCD detector controller configurations are anticipated to be less homogeneous than CMOS detector configurations. Depending on the application, the required number of video channels may vary from four to sixteen and constraints regarding clock control may require more than one clock module per detector. Due to the modularity of NGCII all those requirements can be accommodated. Table \ref{tab:dual_CCD} shows a high-density configuration for one or two simple CCD detectors, e.g. the CCD231. In case of more complex CCD detectors only one detector per shelf is feasible. The empty slots 4 to 6 can be used to populate additional video or clock boards.
\begin{table}[ht]
\caption{Dual CCD Detector} 
\label{tab:dual_CCD}
\small
\begin{center}
\begin{tabular}{|l|l|l|l|l|l|} %% this creates two columns
\hline
\rule[-1ex]{0pt}{3.5ex} Slot & AMC Module & Zone 3 & RTM Module & Det. & Comment \\ \hline \hline
\rule[-1ex]{0pt}{3.5ex} 1 & IOxOS IFC1414 & Digital LVDS & CCD C24 RTM & A & \\ \hline 
\rule[-1ex]{0pt}{3.5ex} 2 & AIES MFMC & Digital LVDS & CCD B24 RTM & A &\\ \hline 
\rule[-1ex]{0pt}{3.5ex} 3 & AQ22 AMC & Analog & CD AQ8 RTM & A &\\ \hline 
\rule[-1ex]{0pt}{3.5ex} 4 & AIES MFMC & Digital LVDS & CCD C24 RTM & B & Optional \\ \hline 
\rule[-1ex]{0pt}{3.5ex} 5 & AIES MFMC & Digital LVDS & CCD B24 RTM & B & Optional\\ \hline 
\rule[-1ex]{0pt}{3.5ex} 6 & AQ22 AMC & Analog & CD AQ8 RTM & B & Optional\\ \hline 
\end{tabular}
\end{center}
\end{table} 

%-----------------------------------
%            Interfaces 
%-----------------------------------
\section{Interfaces}
The interfaces of the detector control hardware connect the detector controller to the rest of the detector system as well as ELT/VLT infrastructure. 

Regarding communication interfaces, the ALICE and LISA wavefront cameras\cite{Marchetti2022} have set the ESO internal standard for detector readout based on ELT infrastructure requirements. NGCII interfaces are designed to be compatible with the wavefront sensor cameras.

\subsection{Real Time Image Network Interface}
The real-time image network is a dedicated network in the ELT infrastructure for transmitting image data in a RTMS/MUDPI\cite{Suarez_2023} multicast stream. As with the wavefront sensor cameras ALICE and LISA\cite{Marchetti2022} the real time network interface is implemented as a 10GBASE-LR optical Ethernet interface.

\subsection{Command and Control Network Interface}
The command and control network is a non-deterministic network for controlling the detector control hardware as well as receiving telemetry data. The interface is implemented as a 1GbE optical or copper Ethernet interface. 

\subsection{PTP Time Reference Network}
In the ELT the precision time protocol (PTP), IEEE1588 is used for distributing timestamps through the telescope. NGCII is able to receive and process PTP time stamps. PTP timestamps are received either on a dedicated network interface or, alternatively over the command and control network interface.

\subsection{Detector Interface}
\begin{figure}[ht]
\centering
\includegraphics[width=0.7\textwidth]{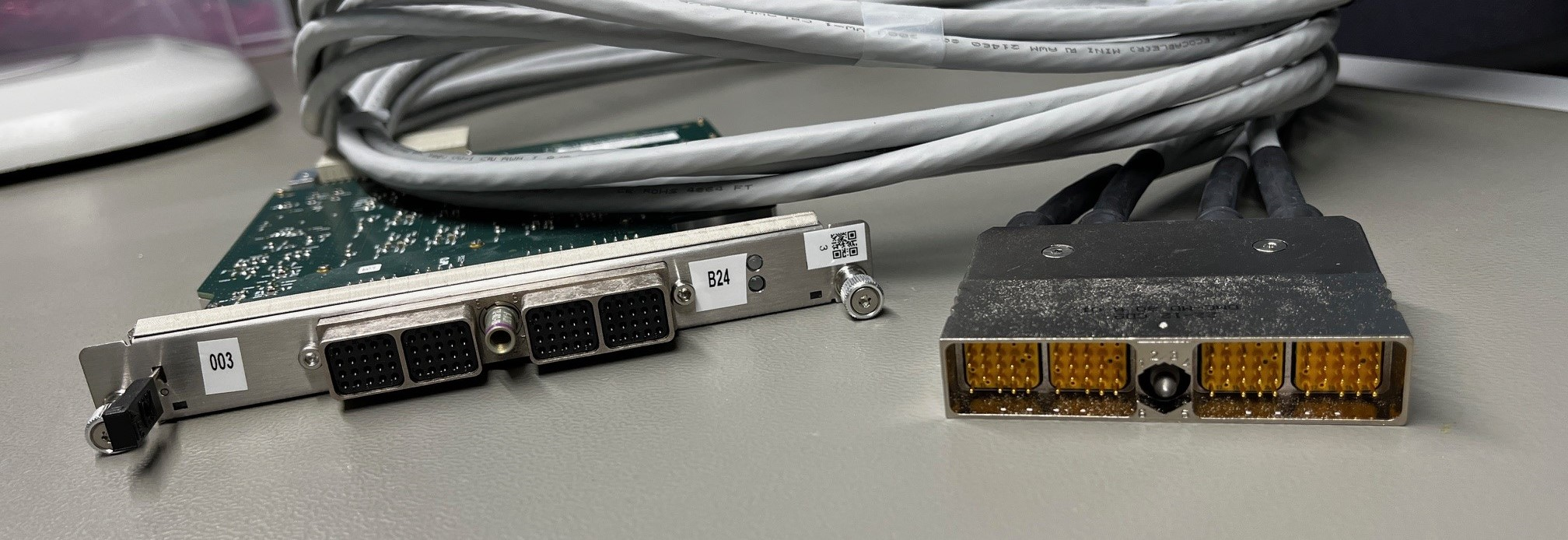}
\caption{Picture of EN4165 Connector System}
\label{pic:en4165}
\end{figure}

The detector interface carries analog bias, clock and video signals from the detector control hardware to the detector and pre-amplifier. On the detector side the interface is typically based on a 128pin MIL connector while on NGC the controller side uses D-SUB connectors. For NGCII the design goal was to use more mechanically robust connectors with sufficient ground pins and mechanical keying that prevents mis-connection of incompatible modules. The main challenge was the limited rear-panel space available on the RTMs that interface the detector. From a shortlist of candidates containing various variants of D-SUB and Micro-D, as well as automotive and military connector systems, the EN4165 system was selected. It is highly configurable, mechanically keyed and has a reliable supply chain with at least two suppliers\cite{amp_en4165,te_en4165} providing inter operable parts. Figure \ref{pic:en4165} shows the EN4165 connector on the CCD B24 RTM module and an example for a mating cable-side connector.

%-----------------------------------
%    Detector Control Circuits 
%-----------------------------------

\section{Detector Control Circuits}
This article is concerned with conceptual design of NGCII, not the circuit design of individual detector control circuits. Nonetheless some desired improvements for individual circuits, identified during requirements collection, are described here since they inform conceptual design choices as the allocation of circuits to modules.

\subsection{Bias Circuits}
Bias circuits are used to supply DC power to a detector. In the terminology of ESO detector controllers all DC voltages supplied to the detector are referred to as biases although they may be referred to as power supplies and/or biases in the detector data sheet.

The analysis of user requirements shows that different detector types have varying requirements regarding their bias voltage range. Typical CMOS detector biases are in a range from $0V$ to $5.0V$ while CCDs range from $0V$ to $33V$ or $0V$ to $-33V$ depending on detector technology. On NGC, biases for all detector types were generated by the same circuit in different population variants. On NGCII it was decided to develop separate bias circuits for CMOS and CCD type detectors to better optimize them for detector requirements. For system design, a direct implication is that separate bias modules need to be developed but also that the relatively simple CMOS bias channels can be combined on one module with the CMOS clock circuits, saving shelf space.

As identified in data from the KMOS instrument\cite{George2018} and further analyzed using the CRIRES+ detector systems which both used HAWAII-2RG detectors and \cite{George2018_2} following the PCA by NASA\cite{Rauscher2013}, thermal drift, the change of DC voltage over detector controller temperature, and low frequency noise are critical performance metrics for CMOS detector operation. The technical note on KMOS testing\cite{George2018} recommends RMS noise not to exceed $15\mu V_{RMS}$ and drift not to exceed $15\mu V/K$. This is addressed by using a temperature stabilized voltage reference LTZ1000 \cite{man:ltz1000} combined with extremely low drift matched resistors\cite{LT5400,Vishay2016} and chopper stabilized operational amplifiers. 

Design verification of two CMOS C20B20 RTM prototype modules shows that the requirement is easily met by every single channel in the ambient temperature range from 0\textdegree C to 40\textdegree C and that the mathematical model used to predict drift performance in the design phase shows promise. For statistically significant results and final verification of the applied model, it is required to first produce more modules. While forty bias channels were tested so far, each module has a single voltage reference source which is a significant contributor to drift. Once a sufficiently large number of modules are available, more measurements can be performed, and results published. It is planned to also study the effect of initial circuit burn-in on drift performance to derive operating instructions for the users.

\subsection{Clock Circuits}
Clock circuits are control signals for the detector. A core conflict identified during requirements collection was a divergence in requirements between CCD detector clocks and CMOS detector clocks leading to the development of separate clock modules for CCD and CMOS systems.

CMOS detector clocks are relatively simple signals with a fixed low-level at $0V$ (GND) and a high-level typically at $3.3V$. Since they are typically buffered on the CMOS technology detector ROIC, a fast and clean edge transition is desirable. The challenge for CMOS clocks lies both in the desired speed of up to $10MHz$ square wave as well as potential cable lengths of up to $5m$ of $50\Omega$ transmission line between detector controller and ROIC.

CCD detector clocks requirements are different to CMOS detector clock requirements. A CCD requires both serial register clocks and parallel phase clocks shifting rows of pixels. For both types the high- and low-level voltages required are in the range of $\pm$12V. Serial clocks operate at up to $2MHz$ square wave while parallel phase clocks operate much slower at up to $50kHz$ square wave.  Both serial and parallel clocks interact with different detector capacitance. Specifically for parallel phase clocks, the driver waveform is very important for performance since they interact directly with the stored pixel charge.

Therefore, control over aspects of the CCD clock waveform is required. It is desirable to control the ramp rate of CCD clocks to achieve a clean cross-over of clock phase signals at the detector while some users also indicate the need for tri-level clocks for advanced readout modes. Those needs are addressed by implementing parallel CCD clocks using a direct digital synthesis (DDS) circuit that allows driving arbitrary clock waveforms. The required space and power budget for those complex clock circuits can be afforded since CCD controller configurations feature separate boards for clock and bias circuits.

\subsection{Video Circuits}
For analog video channels the goal is implementing incremental improvements over NGC. Collection of user requirements shows that switchable gain and low-pass filters, configurable during operation, would allow more versatile detector system design. This is implemented on both the CCD and CMOS video modules.

On NGC the CCD video inputs feature an analog clamp-and-sample circuit that makes the use of analog clamp and sample mandatory. For NGCII the analog clamp-and-sample circuit is kept in place, but a DC-coupled bypass path is implemented, allowing alternative acquisition schemes like digital clamp and sample.

\subsection{GEOSNAP Detector Control}
The GEOSNAP detector is a fully digital detector with analog to digital converters (ADC) as well as complex sequencing logic integrated into its read out circuit (ROIC). Therefore it is not possible to describe GEOSNAP detector control circuits in terms of bias, clock and video channel. The GEOSNAP ROIC\cite{geosnapFPA} transmits digital pixel data through eight high speed current mode logic (CML) transceivers with embedded clock operating at up to 1.6Gbps per lane. Due to the flexible architecture of NGCII, these signals can be fed through the RTM and zone 3 connector with minimal circuitry directly to the Zynq Ultrascale+ SoC on a commercial-off-the-shelf AMC module and processed there. Remaining circuits include remote-sensed power supplies for the ROIC and control signals that are generated on the RTM.

\subsection{Local Power Distribution}
As described in chapter \ref{sec:shelf}, the MicroTCA power supply unit provides one primary +12V payload power rail to each slot. While there is also +3.3V management power available, that rail is very low power and only intended to power an FRU EEPROM, temperature sensor and some local LEDs. So, there is the need to locally generate all voltage supplies required on the module. The MicroTCA.4.1 expansion described in chapter \ref{sec:mtca41} provides some limited support for distribution of power over the optional rear-backplane, but the feature set was found to be insufficient and the cost in terms of complexity too high. For NGCII those local power supplies include:
\begin{itemize}
\item{I/O voltage supplies, e.g. +3.3V, +1.8V}
\item{FPGA core voltage supplies, e.g. +1.0V}
\item{Analog supply voltages, e.g. ±5V, ±10V}
\item{Output driver supplies, e.g. +36V, ±18V}
\end{itemize}
Most supplies have an output power requirement of approximately 3W or lower. All power rails are implemented using non-isolated switch mode power supplies with common supply topologies listed in literature and chip manufacturer documentation\cite{powertop}. Since switch mode power supplies are a major risk factor for electromagnetic compatibility and noise performance of the controller, mitigation measures are implemented early in the design.

When choosing the topology for a power converter, topologies with continuous output current like the Buck\cite{powertop} or Cuk\cite{powertop} converter are preferred to minimize coupling of switching transients into adjacent circuits. If this is not possible, for example for the SEPIC\cite{powertop} converter, an additional LC output filter is implemented that forces continuous current into the load. If absolute accuracy of the DC supply level is important the output of the filter is post-regulated using a linear regulator.  Analogous to the output, continuous input current is also desirable. Therefore, the power input of the entire module or the input of a group of power supplies (e.g. the analog power supplies) is also LC low-pass filtered. 

Furthermore, all power supplies are locally filtered at the load. On each integrated circuit one or more capacitors with a value from $100 nF$ to $10 \mu F$ is placed together with some local bulk capacitance. The entire circuit is then decoupled from the power supply through an in line $1 \Omega$ to $10 \Omega$ resistor or ferrite bead depending on the application and load current. This helps localizing emissions and preventing cross coupling of noise through the power supply rail.

Regarding switching frequency, relatively high values from $500kHz$ to $2MHz$ are chosen. This is to simplify the design of filters due to smaller values required for capacitors and inductors. Switching frequencies of different converters on the same module are spaced at least $100kHz$ apart if possible, to avoid beat frequencies and make it easier to identify noise sources.

%----------------------------------
%            Firmware 
%----------------------------------

\section{Firmware}
The firmware running on the detector control hardware is a new development due to the large hardware differences between NGC and NGCII, specifically the backplane interconnect that changed from a custom, synchronous bus to packet-based PCIe. The two leading principles for NGCII firmware development are:
\begin{description}
\item[Abstraction]{providing internal interfaces and compartmentalization of functionality so that the system remains maintainable for 10 years and longer and so that new features or bug-fixes in one part of the system have minimal impact on the rest of the system.}
\item[Reconfiguration]{of systems. Hardware modules can be used in a similar function in multiple systems. Firmware is designed in such way that no special firmware images are required for any system configuration and modules can be swapped without requiring firmware update or hardware jumper settings, simplifying spare parts management.}
\end{description}

Within NGCII, custom firmware is exclusively run on AMC modules. The MCH and other MicroTCA components do run firmware which is fully developed and maintained by the module's manufacturer while rear transition modules do not contain programmable hardware. All firmware is upgradeable remotely during operation not requiring physical access to the NGCII unit. Any firmware maintained by ESO is built by a traceable continuous integration pipeline on dedicated build servers and the exact build version of each firmware image can be read through software.

\begin{figure}[ht]
\centering
\includegraphics[width=\textwidth]{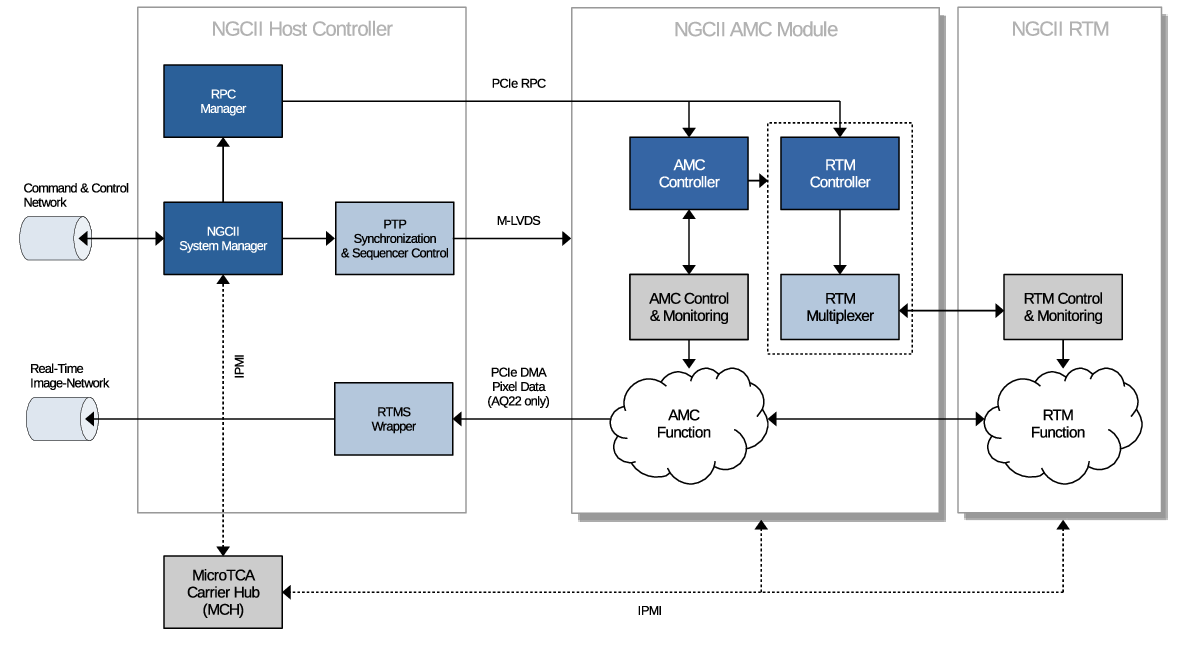}
\caption{NGCII Firmware Architecture}
\label{pic:firmware}
\end{figure}

A sketch of the NGCII firmware architecture depicting overall data flow is shown in Figure \ref{pic:firmware}. 
The core firmware entity is the NGCII system manager running on the host controller module. It communicates through the command and control network and orchestrates all other firmware entities. It has access to the shelf-internal IPMI bus through the MCH and is able to monitor the system status and acquire low-level telemetry data from voltage and temperature sensor.

Modules are controlled through PCIe remote procedure call (RPC) via the RPC manager. Each module runs an AMC controller entity that controls functionality local to the AMC module. Note that the host controller, despite its special role in the firmware architecture, also acts as a regular AMC and features an AMC controller. Since RTMs do not run firmware themselves, they are controlled through an RTM controller running on their associated AMC. To abstract the functionality of the RTM with its various LVDS user I/O lines on the zone 3 connector from the AMC firmware image, a configurable multiplexer is implemented on the AMC to dynamically map RTM controller functions to zone 3 I/O.

Modules that generate pixel data (AQ22 AMC) send these data packets to the host controller via PCIe direct memory access (DMA) which wraps pixel data and enriching information into RTMS frames and transmits them over the real-time image-network.

The system manager also orchestrates intra-shelf synchronization of modules by distributing clock and trigger signals over the M-LVDS bus signals on backplane ports 17 to 20.

%----------------------------------
%        First Light Testing 
%----------------------------------
\section{First-Light Testing}
Bench tests for compliance of NGCII with electrical requirements are ongoing. In parallel, tests with detector ROICs and cold detectors are performed in the ESO internal test facilities comparing performance of NGCII with NGC. There is only very limited system level testing performed inside the NGCII development project, mostly to prove general functionality of NGCII. Detailed detector characterization yielding data for detector system performance with NGCII is to be done in the instrument projects, primarily METIS, HARMONI and MICADO. So far, laboratory first-light has been achieved with a SAPHIRA ROIC and a HAWAII-2RG detector.

\subsection{First Light with SAPHIRA ROIC}
\begin{figure}[ht]
\centering
\includegraphics[width=\textwidth]{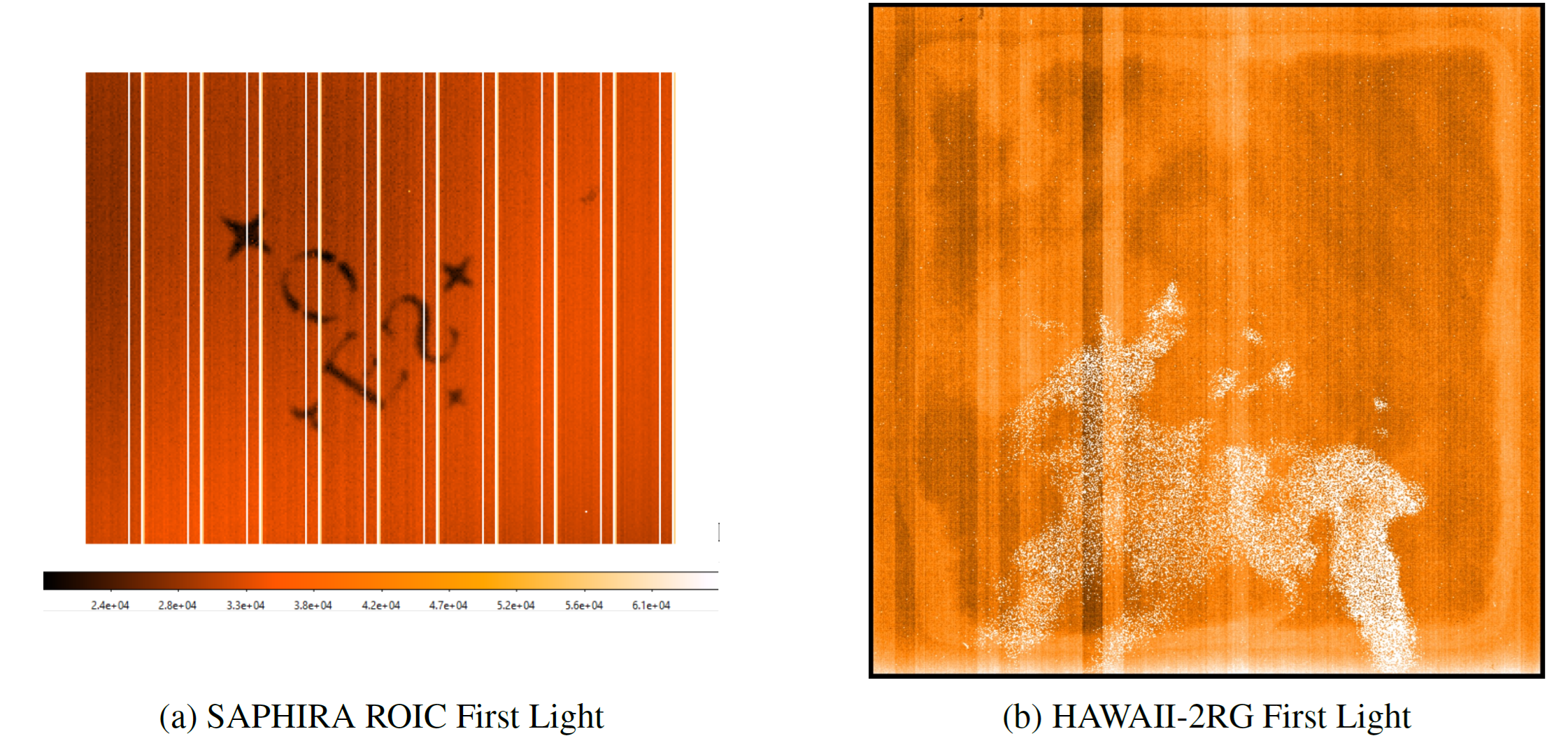}
\caption{CMOS Detector First Light Images}
\label{pic:first_light}
\end{figure}

The very first image taken with NGCII in December 2022 is shown in figure \ref{pic:first_light}a. A non IR-sensitive bare ROIC of a SAPHIRA detector was connected to NGCII using a preamp and exposed to visible light through an ESO-logo stencil. Some video outputs of the setup are known to be non-operational, leading to recurring dead pixel columns.

This test shows that NGCII is operational and that voltage configuration files as well as sequencer programs from NGC are transferable to NGCII although no conclusions about detector performance can be drawn from the results. Further characterization of the SAPHIRA detector with NGCII is planned in the context of the METIS project and to be reported separately.

\subsection{First Tests with HAWAII-2RG Detector in Fast Mode}
Running the HAWAII-2RG detector in fast mode at up to $30fps$ was a challenging task with NGC due to bandwidth limitations and, as listed in section \ref{sec:goal}, more reliable operation at fast speeds was a development goal for NGCII. A HAWAII-2RG engineering detector was placed in a cryostat with a cryogenic, OPAx354 based, preamplifier designed for the METIS project and a $3.5m$ long warm cable. The first full-frame picture taken with the HAWAII-2RG detector is shown in figure  \ref{pic:first_light}b.

Again, it was possible to transfer voltage configuration files and sequencer files from NGC and operate the detector.

\begin{figure}[ht]
\centering
\includegraphics[width=0.9\textwidth]{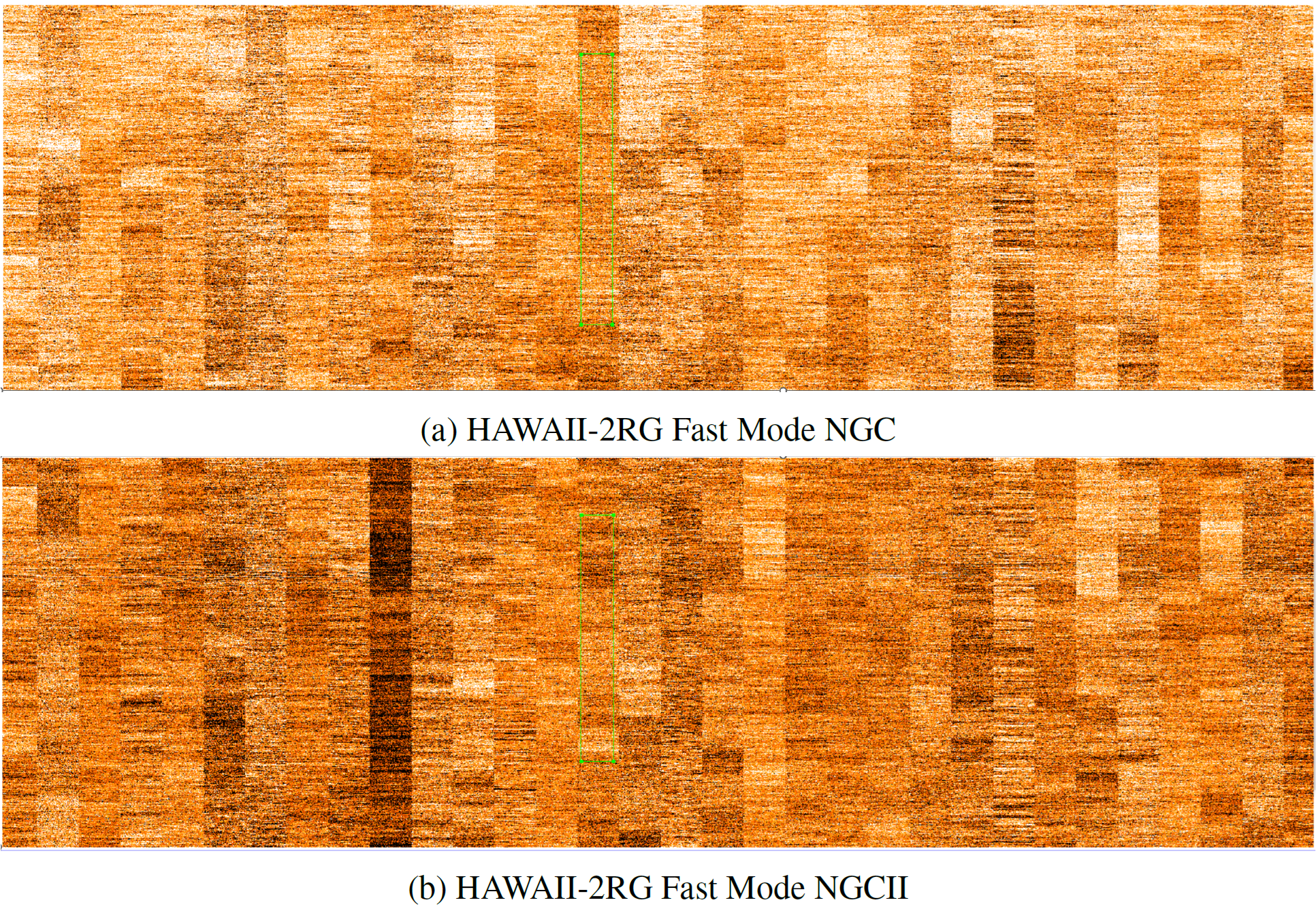}
\caption{HAWAII-2RG Fast Mode Difference Bias Image with NGC and NGCII}
\label{pic:diff_image}
\end{figure}

\begin{figure}[ht]
\centering
\includegraphics[width=\textwidth]{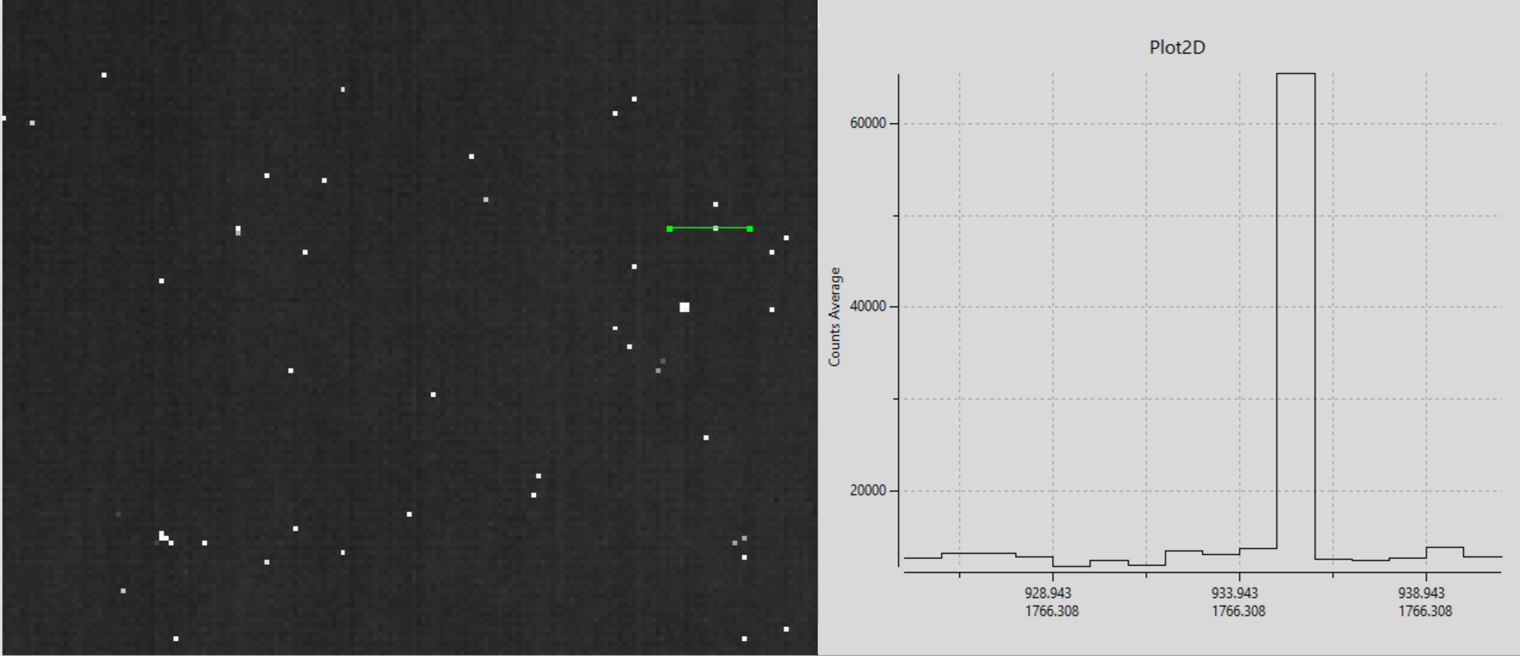}
\caption{Pixel Step Response}
\label{pic:step_response}
\end{figure}

Since the connector interface at the cryostat wall is fully compatible for NGC and NGCII, some basic measurements were performed to compare the two detector controllers in the same setup. Difference bias images of the HAWAII-2RG detector with NGC and NGCII are shown in figure \ref{pic:diff_image}.
The patterns visible in both images are changing over time and are generally understood to be $1/f$ noise of the OPAx354 operational amplifier on the preamp. Visually both images are very similar. Read noise and conversion gain are measured with both NGC and NGCII. Measured values in table \ref{tab:det_params} are close to each other and it is concluded that NGCII performs similarly to NGC regarding gain and noise.
\begin{table}[ht]
\caption{HAWAII-2RG Detector Parameters} 
\label{tab:det_params}
\small
\begin{center}
\begin{tabular}{|l|l|l|l|} %% this creates two columns
\hline
\rule[-1ex]{0pt}{3.5ex} Parameter & NGC Measured & NGCII Measured &  Unit \\ \hline \hline
\rule[-1ex]{0pt}{3.5ex} Read Noise & $22.3$ & $20$  & $e-_{RMS}$ \\ \hline 
\rule[-1ex]{0pt}{3.5ex} Conversion Gain & $3.73$ & $3.89$ & $\mu V/e-$\\ \hline 
\end{tabular}
\end{center}
\end{table} 
The pixel step response of a hot pixel with $240ns$ pixel period (corresponding to approximately $30fps$) is also measured and shown in figure \ref{pic:step_response}. Since there is no visible trailing of the hot pixel the preliminary conclusion is that NGCII runs well at $30fps$ with a HAWAII-2RG detector. Final performance is to be evaluated during detector testing in the METIS project.
As an experiment the pixel period was reduced to $80ns$ (approximately $90fps$). The resulting image is not meaningful since the preamp severely limits the bandwidth of the system but NGCII reliably transmits images over the real-time network at a approximate data rate of $6Gbps$.
%----------------------------------
%             Discussion
%----------------------------------

\section{Discussion}
A system architecture for the future ESO detector controller NGCII has been defined. Due to the early decision to base the detector controller on a pre-existing modular platform it was possible to focus development efforts on detector specific circuits. Laboratory first-light was achieved, concluding the system design presented in this work.

\subsection{Future Work}
This work presents a snapshot in time of NGCII development. Until integration of the detector controller into the ELT is possible, several more steps need to be completed.

\subsubsection{First-light with further Detectors}
After first light with HAWAII-2RG and SAPHIRA detectors has been achieved and the system concept proven it is required to continue working on completing detector control for all detectors required for the first milestone listed in Table \ref{tab:detectors1}. This leaves:
\begin{itemize}
\item{CCD Detectors}
\item{GEOSNAP Detector}
\end{itemize}
Efforts are ongoing in parallel with steps to prepare mass production of CMOS systems.

\subsubsection{System Level Testing}
For all supported detector types it is required to run a set of tests fully characterizing the entire system. While the final performance of NGCII can only be validated on-sky, this will shorten integration time of the controller with the detector system and instrument. Due to the internal project structure and available resources, detailed detector testing will be performed in the context of ESO internal detector testing for the instruments.

\subsubsection{Environmental Compliance Testing}
Due to the complexity of the controller a test campaign is planned to get results on environmental compliance as soon as possible. Representative configurations of NGCII will be tested over the full temperature range that the controller is required to operate in. Furthermore pre-compliance tests will be performed to evaluate electromagnetic compatibility of the controller with representative auxiliary equipment at a test site. This is to speed up integration and instrument level compliance test by eliminating potential issues early.

\subsubsection{Preparation of Series Production}
With general functionality proven and first tests showing encouraging performance figures, series production is being prepared. This includes fixing minor hardware issues discovered during testing, improving testability and manufacturability of ESO-design modules and placing orders for commercial components. 

\subsubsection{Further Results on Circuit Performance}
This work describes the high level system design of NGCII and does not go into detailed design and characterization of the system. Many design calculations, simulations and tests of detector control circuits, for example extremely low drift bias circuits and arbitrary clock waveform generation for CCD clocks, were performed. These results will also be presented at a later date to the community.

NGCII is used for ESO-internal detector integration and testing on a daily basis. Those results should also be compiled and published, preferably in the context of the instrument they were tested for.

\subsection{Further Development of MicroTCA.4}
The MicroTCA.4 standard is under active development. NGCII detector controller development is closely tied to changes in the MicroTCA standard and needs to react to opportunities or changes in the ecosystem. PICMG discourages changes that break backwards compatibility so in principle staying on the base standard MicroTCA.4 indefinitely is an option.

\subsubsection{MicroTCA 4.1}
\label{sec:mtca41}
Enhancements to MicroTCA.4 are standardized in an additional PICMG standard MicroTCA.4.1\cite{mtca41}. Besides codification of RTM application classes as described in section \ref{sec:rtm}, it introduces:
\begin{itemize}
\item{Auxiliary Backplane for Rear Transition Modules (RTM)}
\item{MCH Management Support \& Extended Rear Transition Module (MCH-RTM)}
\item{Rear Power Modules (RPMs)}
\end{itemize}
None of the above features are required for NGCII so MicroTCA.4.1 is not implemented.

\subsubsection{Next Generation MicroTCA}
Further enhancements are in development\cite{Rehlich2023} with the goal to standardize in an extension of MicroTCA.4. The current drafts of the new enhancements include features that may become interesting for future iterations of NGCII.
\begin{description}
\item[Increased Backplane Lane Speed]{It is planned to upgrade the PCIe profile of the fat-pipes from PCIe Gen.3 to PCIe Gen.5 approximately quadrupling backplane bandwidth with the same number of PCIe lanes.}
\item[Increased Modules Power]{Module power is set to increase from 80W per AMC/RTM pair to 110W. This includes the additional cooling required for dissipating the heat generated.}
\end{description}
For larger and faster detectors these developments in the MicroTCA ecosystem shall be monitored and upgrades to NGCII implementation made where necessary.

\section{Conclusion}
The architecture for the new ESO modular detector controller (NGCII) was defined after the decision to develop it in house to meet ESOs operational and infrastructure requirements. Laboratory first light was achieved for SAPHIRA and HAWAII-2RG detectors with promising first performance results. With focus placed on operational reliability and system stability very early in the development process, for example by choosing a base system standard (MicroTCA.4) with a proven track record in the particle physics community, or the modular firmware architecture, it is anticipated that the final systems delivered will form a stable base of large astronomical instruments for the ELT era.

\subsection* {Code, Data, and Materials Availability} 
The data that support the findings of this study are available from the corresponding author upon reasonable request.

%%%%%%%%%%%%%%%%%%%%%%%%%%%%%%%%%%%%%
%%%%%%%%%%%    REST OF PAPER   %%%%%%%%%%%%%%%%%
%%%%%%%%%%%%%%%%%%%%%%%%%%%%%%%%%%%%%

\bibliography{article} % bibliography data in report.bib
\bibliographystyle{spiejour} % makes bibtex use spiejour.bst

%%%%% Biographies of authors %%%%%

\vspace{2ex}\noindent    \textbf{Mathias Richerzhagen} is a detector electronics engineer at the European Southern Observatory. He received his Engineering Diploma from RWTH Aachen University in 2012. His current research includes development of the detector controller for the ELT as well as some work in cryo-electronics.

\vspace{1ex}
\noindent Biographies and photographs of the other authors are not available.

\listoffigures
\listoftables

\end{spacing}
\end{document}